\documentclass[12pt,preprint]{aastex}

\slugcomment{Not to appear in Nonlearned J., 45.}

\shorttitle{SEPARATING LINERS INTO TWO POPULATIONS}
\shortauthors{Wang et al}

\begin{document}


\title{TWO PHYSICALLY DISTINCT POPULATIONS OF LOW-IONIZATION NUCLEAR EMISSION-LINE REGIONS
}


\author{J. Wang\altaffilmark{1}, J. Y. Wei\altaffilmark{1}, and P. F. Xiao\altaffilmark{1}}
\affil{National Astronomical Observatories, Chinese Academy of Science, Datun 20A, 
Chaoyang District, Beijing, 100012}
\email{wj@bao.ac.cn}



\begin{abstract}

The nature of Low-ionization Nuclear Emission-line Regions (LINERs) has been an 
open question for a long time. We study the properties of LINERs from several 
different aspects. The LINERs are found to consist of two different categories
that can be clearly separated in the traditional BPT diagrams, especially in 
the [\ion{O}{1}]/H$\alpha$ vs. [\ion{O}{3}]/H$\beta$ diagram. LINERs with 
high [\ion{O}{1}]/H$\alpha$ ratios (population I) differ from 
ones with low ratios (population II) in several properties. Broad 
emission lines are only identified in the spectra of population I LINERs. 
While only the population II LINERs show luminous infrared emission and 
occurrence of core-collapse supernovae in the host. Combining these 
results with the known distribution of stellar populations not only suggests 
that the two populations have different line excitation mechanisms, 
but also implies that they are at different evolutionary stages.

\end{abstract}





\section{INTRODUCTION}

LINERs (Low Ionization Nuclear Emission-line Region) are
frequently identified in the local universe:  different surveys show that
up to 1/3 of all galaxies show LINER-like spectra (see Ho 2008 for a recent
review). Comparing with Seyfert 2 galaxies and star-forming
galaxies, optical spectra of LINERs are dominated by the emission lines of
low-ionized species, such as [\ion{O}{1}]$\lambda6300$,
[\ion{O}{2}]$\lambda3727$, [\ion{S}{2}]$\lambda\lambda6716,6731$ and
[\ion{N}{2}]$\lambda6548,6583$ (e.g., Heckman 1980) 

Emission-line ratios are frequently used to classify different emission-line 
galaxies with distinct power sources. At present, the classification 
of emission-line galaxies is commonly based on the empirical BPT diagrams 
that were originally proposed by Baldwin et al. (1981), and then refined by 
Veilleux \& Osterbrock (1987). Theoretical and empirical 
demarcation lines were proposed by Kewley et al. (2001) and Kauffmann et al. (2003) 
to separate star-forming galaxies on the BPT diagrams, respectively.
Kewley et al. (2006) recently proposed a new empirical
classification scheme that separates Seyferts and LINERs. 

However, LINERs are diverse in their observational properties.
Ho et al. (1993) identified a group of LINER/\ion{H}{2} ``transition objects'' with relatively low 
[\ion{O}{1}]/H$\alpha$ ratios compared with ``pure'' LINERs. They proposed 
that the LINER/\ion{H}{2} objects are most likely composite systems that contain 
a LINER plus an \ion{H}{2} region. Many studies found that  
these ``transition'' LINERs differ from the ``pure'' ones in their stellar 
population distributions (e.g., Cid Fernandes et al. 2004; Kewley et al. 2006).
Wang \& Wei (2008) and Xiao et al. (in preparation) show that all their broad-line LINERs 
are associated with large [\ion{O}{1}]/H$\alpha$ ratios and old stellar populations. 

LINERs have been under hot debate in origin for a long time.
The photoionization by the central
AGNs is suggested to be a main power source in some LINERs (e.g., Ferland \& Netzer 1983).
The AGN origin is strongly supported by the
detection of broad emission lines in optical spectra (e.g., Ho et al. 1997; Wang \& Wei 2008).
Other evidence supporting the AGN origin includes the hard X-ray spectra 
(e.g., Terashima et al. 2000; Flohic et al. 2006),
and UV variability (e.g., Maoz et al. 2005). On the other hand, 
LINERs could be explained by other 
excitation mechanisms unrelated to an AGN, such as fast shocks (e.g., Dopita et al. 1997; 
Lipari et al. 2004), post-AGB stars (Binette et al. 1994),
or low mass X-ray binaries (e.g., Eracleous et al. 2002; Flohic et al. 2006).

At present, these results imply that the current classification scheme results in a hybrid LINER 
population (different power sources or evolution stages). 
In this letter, we study the classification scheme through several different approaches, and 
show that the currently classified LINERs consist of two populations
that can be clearly separated on the BPT diagrams. The distinct
two populations might have different physical origins, and be at different evolutionary stages.

\section{BROAD-LINE LINERS: A ZONE OF AVOIDANCE IN THE BPT DIAGRAMS}

Wang \& Wei (2008) and Xiao et al. (in preparation)\footnote{Xiao et al. searched 
for partially obscured AGNs from the galaxies (the median S/N ratio per pixel $\geq30$)
located above the Kauffmann demarcation line according to measurements given in the 
MPA/JHU catalog by the same method described in Wang \& Wei (2008). 
}
systematically searched for 
partially obscured AGNs from the MPA/JHU SDSS DR4 catalog
(Heckman \& Kauffmann 2006 and references therein) according to the existence 
of broad H$\alpha$ emission in the spectra\footnote{The partially obscured AGNs are
selected if $F_w/\sigma_c\geq3$, where $F_w$ is the stellar continuum subtracted 
flux averaged over the rest frame
wavelength range from 6500 to 6530\AA, and $\sigma_c$ is the standard deviation of flux
within wavelength region between 5980 and 6020\AA.}. 
Figure 1 shows the distributions of these partially obscured AGNs on the three 
commonly used BPT diagrams. 
The objects from Wang \& Wei (2008) and from 
Xiao et al. are displayed by the red solid and open squares, 
respectively. Indeed, the detection of broad H$\alpha$ emission is strongly 
affected by the contamination of diffuse light from the old stars.
A large effect is expected in the SDSS spectra, especially 
for weak nuclear emission, since the spectra are obtained by a relatively 
large fiber aperture (3\symbol{125}).
The nearby broad-line LINERs identified in 
the Palomar spectroscopic surveys (Ho et al. 1997) are therefore included 
to avoid the bias. 
These objects are shown in Figure 1 by the blue circles.

Based on the combined sample, although it is not clear in the [\ion{O}{3}]/H$\beta$ vs. 
[\ion{N}{2}]/H$\alpha$ diagram, a remarkable zone of avoidance (ZOA) is obviously found in the 
other two diagrams, especially in the [\ion{O}{3}]/H$\beta$ vs. [\ion{O}{1}]/H$\alpha$ diagram,
for the broad-line LINERs. The non-detection of broad H$\alpha$ in the 
objects within the ZOA can not be caused by the increasing contribution 
of young stars, both because young stars show less contamination caused by the stellar absorptions 
at the H$\alpha$ region and because Seyfert galaxies have stellar population ages younger than 
LINERs (see Figure 11 in Kewley et al. 2006). We examined the high quality spectra of LIRGs (Cao et al. 2006)
located in the ZOA one by one by eyes, and the results reinforce the non-detection of broad
H$\alpha$.
The ZOA naturally suggests that the previously 
defined LINERs likely consist of at least two types of galaxies with distinct power sources. 
In each panel, we define an empirical lower boundary for these broad-line LINERs by eye, and 
plot the boundary by a short-dashed line. 
The regions above and below the line is defined as Zone A 
and B for short, respectively.





\section{WHAT IS IN THE ZOA?}

The lack of broad-line LINERs naturally motivates
us to suspect that the LINERs in the Zone B are powered by the ionizing radiation
arising from some stellar processes unrelated to an AGN. 
The host properties are investigated in two aspects in this section: infrared emission and 
occurrence of core-collapse supernovae (cc-SNe).

\subsection{Ultra-luminous/luminous Infrared LINERs}

Cao et al. (2006) 
established a large sample of luminous infrared galaxies 
(LIRGs, defined as $L_{\mathrm{IR}}>10^{11}L_\odot$) through the cross-identification 
of the SDSS DR2 with the catalogs of the \it IRAS \rm survey.
Figure 2 shows the distributions of the sample on the three BPT diagrams. 
The LIRGs are shown by the cyan solid 
and open triangles for the \it IRAS \rm faint and point sources, respectively.  
As an additional test, the ultra-luminous infrared galaxies 
(ULIRGs, $L_{\mathrm{IR}}>10^{12}L_\odot$) identified 
in the \it IRAS \rm 1Jy sample (Kim \& Sanders 1998; Veilleux et al. 1999) are overplotted
in Figure 2 by the blue stars. 

Similar as done in Sect. 2, we focus on the [\ion{O}{3}]/H$\beta$ vs.
[\ion{S}{2}]/H$\alpha$ and [\ion{O}{3}]/H$\beta$ vs. [\ion{O}{1}]/H$\alpha$ diagrams.
The diagrams clearly show that all the infrared luminous LINERs are remarkably
clustered in the Zone B defined previously, except only a few outliers.

\subsection{LINERs: Occurrence Of cc-SNe}

The occurrence of cc-SNe is an indicative of young and intermediate-aged 
stellar populations. The cc-SNe are believed
to be produced by the explosion of massive stars ($\geq8-10M_\odot$) at the
end of their life time ($\sim10^{7.5}$yr).
We cross-match the SDSS DR6 with the SAI-SDSS image SN catalog originally 
done by Prieto et al. (2008). The sample finally contains the SDSS spectra of host 
galaxies (central 3\symbol{125}) of 51 type Ib/c and 182 type II SNe. 
We remove the absorption features from the spectra 
through the PCA technique as described in Wang \& Wei (2008). 
The emission line fluxes are then measured by the direct integration on the starlight-subtracted 
spectra. The distributions of these galaxies on the three BPT diagrams are plotted in Figure 2. 
As same as the results given before, the LINERs selected by the occurrence of cc-SNe are mainly 
located in the Zone B.

\section{COMPARISON BETWEEN THE TWO POPULATIONS OF LINERS}

The above statistical study indicates that 
the currently classified LINERs could be obviously 
separated into two categories on the traditional BPT diagrams. 
The separation is 
most remarkable in the [\ion{O}{1}]/H$\alpha$ vs. [\ion{O}{3}]/H$\beta$ diagram:   
a) population I LINERs within the Zone A have large [\ion{O}{1}]/H$\alpha$ ratios; 
b) population II LINERs within the Zone B have low [\ion{O}{1}]/H$\alpha$ ratios. 
The two populations show distinct observational properties that are
compared with each other in Table 1. In addition to the observational properties 
described above, the two populations are also different in their host 
stellar population ages. The stellar population is found to be on average younger in
population II LINERs than in population I LINERs
(e.g., Kewley et al. 2006; Wang \& Wei 2008; 
Cid Fernandes et al. 2004; Gonzalez Delgado et al. 2004; 
Sarzi et al. 2005)\footnote{The over detected young stellar population in the ``transition'' 
LINERs are recently argued against by Ho (2008). The author believes that the over 
detection rate might be caused by the Hubble type bias.}.


The distinct observational properties imply that the two population LINERs have 
different excitation mechanisms. 
For the population I LINERs, one naturally expects that the gas is 
photoionized by the ionizing radiation arising from the central 
accretion disk due to the detection of the broad lines in their spectra. Not as 
luminous AGNs, a weak ionizing radiation ($\log U\sim -3- -4$) with a hard 
spectrum is required to explain the emission-line spectra of LINERs 
(e.g., Kewley et al. 2006; Lewis et al. 2003; Sabra et al. 2003; Gabel et al. 2000).
Indeed, the hard X-ray spectra are observed in some LINERs by different instruments 
(e.g., Flochi et al. 2006; Gliozzi et al. 2008; Rinn et al., 2005).
The low level and hard ionizing radiation is consistent with the predictions of 
the advection dominated accretion model (e.g., Abramowicz et al. 1995; Narayan et al. 1995)
in which the inefficient radiation results in a hot accretion disk. 
Finally, the non-detection of luminous infrared emission 
and cc-SNe in the objects could be interpreted by both weak accretion activity 
and aged stellar population.

The explanation of the excitation mechanism is, however, non-trivial for the population II 
LINERs. The non-detection of broad lines in their spectra implies that they
are plausibly related to the stellar processes (see also in Ho 2008). 
Comparisons of X-ray and optical observations show an anomalous high H$\alpha$ emission  
for the ``transition'' LINERs (e.g., Ho 2008; Terashima \& Wilson 2003; Flochi et al. 2006).
The fast shock model could be excluded at first because the observed UV high excitation 
emission lines are always lower than the predicted ones (e.g., Gabel et al. 2000; Maoz et al. 1998).
The young star cluster model (Barth \& shields 2000) requires extremely young stellar populations and sizable
Wolf-Rayet stars that are, however, not observed (e.g., Ho et al. 2003).
The ``transition'' LINERs are suspected to be composite systems that
consist of a LINER and an \ion{H}{2} region component, which, however,
can not explain the non-detection of broad emission line in the population II LINERs.

The LINER-like spectra can be ionized by evolved, low massive stars. 
The spectral synthesis model shows 
that the ionizing photon is dominated by the emission from the post-AGBs after $\sim10^8$yr since 
the burst (Binette et al. 1994). A simple calculation recently done by Ho (2008) suggests 
that the post-AGB stars within the center of galaxies can provide sufficient ionizing photons in the 
``transition'' LINERs. In particular, the post-AGB star models
predict a low [\ion{O}{1}]/H$\alpha$ ratio ($<0.5$)
for reasonable parameters (Binette et al. 1994; Stasinska et al. 2008), 
which is quantitatively in agreement with the current boundary separating the population I and II LINERs. 
The post-AGB star model is also consistent with the observed infrared properties. 
One of the characteristic features of post-AGB stars are their strong infrared excess. We estimate 
the infrared luminosity for individual post-AGB star as 
$L_{\mathrm{IR}}^*/L_\odot=1.77f(r/\mathrm{pc})^2 (T_\mathrm{d}/\mathrm{K})^4$ 
in two fiducial cases: $r=0.01$pc, $T_d=200$K and $r=0.1$pc, $T_d=80$K (Suarez et al. 2006), 
where $r$ is the distance from the dust shell to star, $T_\mathrm{d}$ 
is the temperature of the dust and $f=0.004$ is the emissivity of the dust (Draine \& Lee 1984).
The total infrared luminosity from the dust is then $L_{\mathrm{IR}}=L_{\mathrm{IR}}^*(M/M_*)$, 
where $M$ and $M_*$ is the integrated post-AGB star mass
and mass of individual post-AGB star (i.e., $M_*=1-8M_\odot$).   
Adopting the typical integrated stellar mass of galaxy center $\sim2\times10^9 M_\odot$ and the Salpeter
IMF results in a roughly estimated integrated infrared luminosity $\sim10^{10.7-12} L_\odot$, 
which is (marginally) in agreement with the definition of LIRGs (ULIRGs). 
Note that the estimated infrared luminosity should be strictly regarded as an upper limit 
because the exact value strongly depends on star formation history of galaxy.
Finally, the occurrence of cc-SNe in 
the population II LINERs provides evidence in timescale supporting the post-AGB star model.
As massive stars explode at the end of their lifetimes $\sim10^{7.5}$yr, the low- and intermediate-mass 
stars (1-8$M_\odot$) evolve to the AGB and post-AGB phase at the end of their main sequence 
lifetimes with an upper limit $\sim5\times10^7$yr.

\section{IMPLICATIONS ON AGN'S EVOLUTION}

We briefly discuss the implications on the AGN's evolutionary scenario in the context 
of the coevolution of AGN and its host galaxy. 
Some recent studies suggest a delay (with time scale $\sim100$Myr) between the onset of the star formation
and the subsequent intensive black hole accretion 
(e.g., Davis et al. 2007; Kawakatu et al. 2003; Granato et al. 2004). 
In this scenario, the emission lines might be dominantly excited by the massive starburst
at the very beginning of an AGN activity. 
The central accretion nucleus is either obscured (e.g. Hopkins et al. 2005, Wada \& Norman 2002) 
or too weak to maintain a BLR 
(e.g., Laor 2003; Xu \& Cao 2007; Nicastro et al. 2003; Elitzur \& Shlosman 2006). 
The accretion activity might increase within the first a few of $10^8$yr 
as the very young circumnuclear stellar
population ages (Davis et al. 2007). 
After the first a few of $10^8$yr, when the AGN is still buried or weak, 
the emission lines are dominantly excited by the ionizing radiation from post-AGB stars 
(i.e., the population I LINER phase). 
Sturm et al. (2006) indicated that IR-bright LINERs have intermediate PAH feature ratios 
(12.7$\mu$m/11.2$\mu$m and 6.2$\mu$m/11.2$\mu$m) between \ion{H}{2} region and IR-faint LINERs.
At this phase, the post-AGB stars contribute the luminous 
infrared emission, and the evolved massive stars explode at the end of their lives.
A luminous AGN subsequently appears, and shines with decreasing 
Eddington ratio as the stellar population continuously ages (Wang \& Wei 2008;
Kewley et al. 2006). After the luminous phase, 
the AGN evolves to a less luminous stage with hard ionizing radiation (i.e., population I LINER
phase). The underlying very old stellar population becomes dominant because the 
importance of the young stellar population fades out entirely.

\section{SUMMARY}

The previously defined LINERs are found to consist of two categories that can 
be clearly separated in the traditional BPT diagrams, especially in the 
[\ion{O}{1}]/H$\alpha$ vs. [\ion{O}{3}]/H$\beta$ diagram. Population I LINERs
are defined to have higher [\ion{O}{1}]/H$\alpha$ ratios than 
Population II LINERs. The two population LINERs differs in the presence of 
broad emission lines in their spectra, strength of infrared emission, host galaxy stellar population age,
and occurrence of cc-SNe. These different properties suggest 
that the two populations not only have different line excitation mechanisms, 
but also are at different evolution stages.

\acknowledgments
We would like to thank the anonymous referee for valuable comments and 
suggestions that help to improve the paper.
This work is supported by the National Science Foundation of China under 
grants 10503005 and 10803008, and by the National Basic Research Program of China 
(grant 2009CB824800).

\clearpage



\begin{figure}
\epsscale{.80}
\plotone{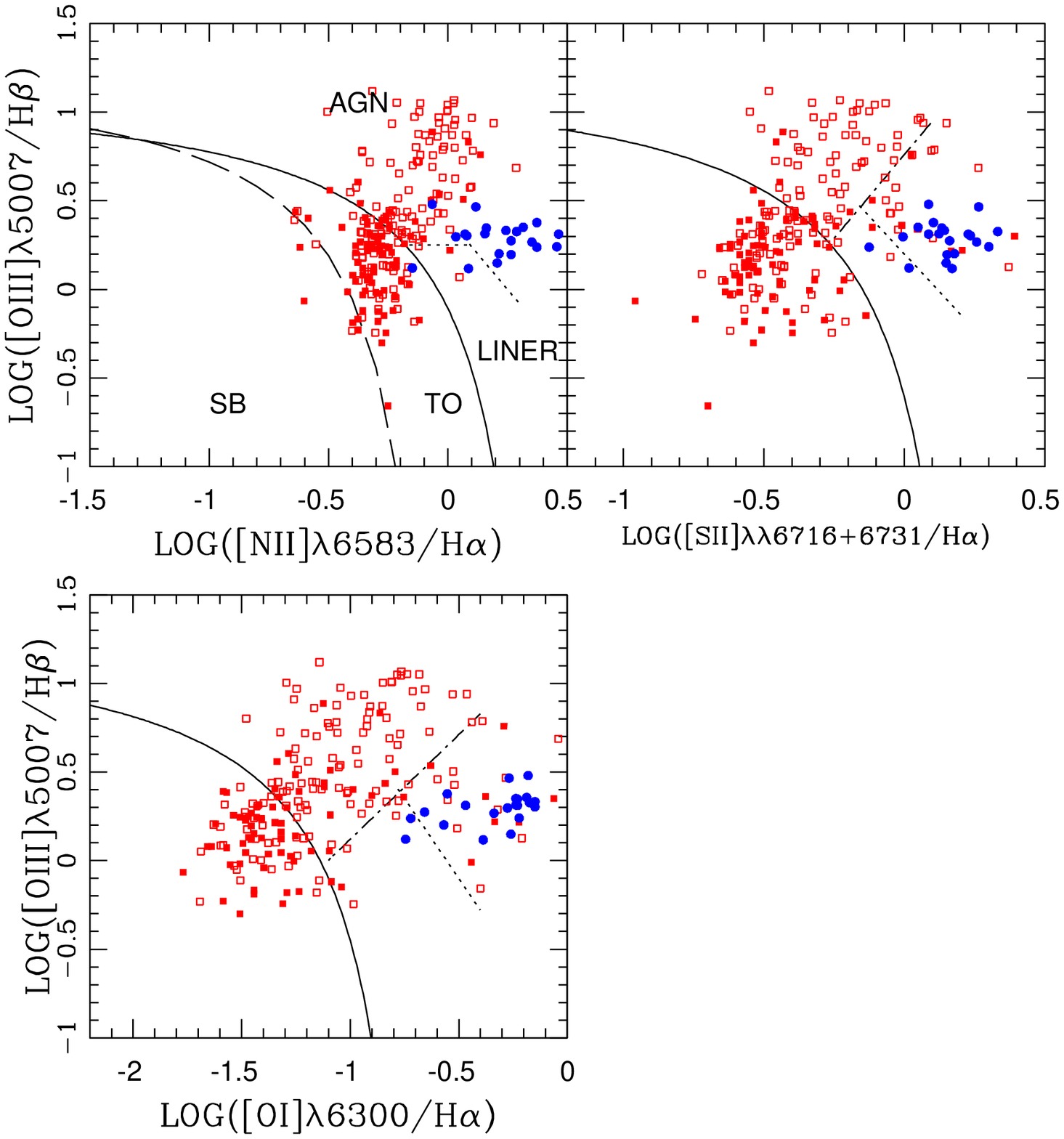}
\caption{The three diagnostic BPT diagrams. The solid lines show the theoretical
demarcation lines separating AGNs from star-forming galaxies proposed by
Kewley et al. (2001), and the long-dashed line the empirical line proposed
in Kauffmann et al. (2003). The empirical separation scheme of LINERs suggested
by Kewley et al. (2006) is drawn by the dot-dashed lines in the
[\ion{S}{2}]/H$\alpha$ vs. [\ion{O}{3}]/H$\beta$ and [\ion{O}{1}]/H$\alpha$ vs. [\ion{O}{3}]/H$\beta$
diagrams. The red squares show the partially obscured AGNs selected from the MPA/JHU
catalog (solid squares represent objects from Wang \& Wei (2008), and open ones from Xiao et al. in
preparation.) The broad-line LINERs identified in the Palomar spectroscopic surveys are plotted by
the blue circles (Ho et al. 1997). The short-dashed lines show the proposed 
boundaries of the broad-line LINERs drawn by eyes. We define the LINERs 
located above the boundaries as population I LINERs, and ones below the boundaries 
as population II LINERs.
}
\end{figure}

\clearpage
\begin{figure}
\epsscale{.80}
\plotone{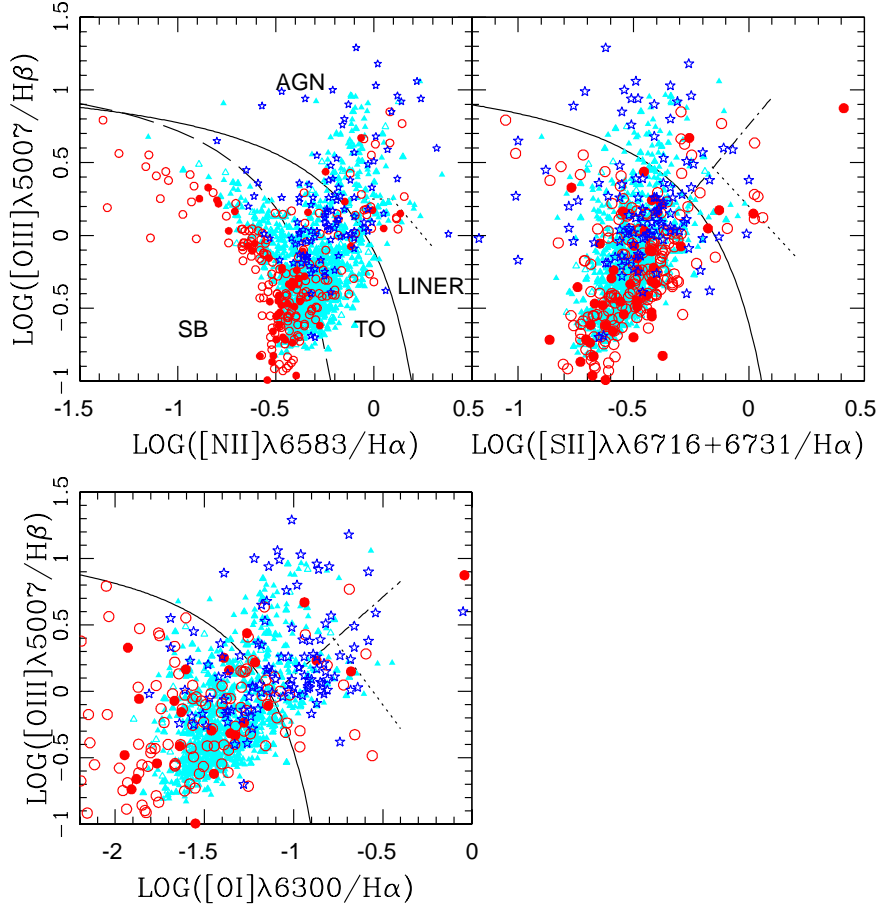}
\caption{The same as Figure 1 but for different samples. The cyan solid and 
open triangles show the LIRGs that are identified by cross-matching the SDSS DR2 with 
the \it IRAS \rm faint and point sources, respectively (Cao et al. 2006). The ULIRGs
from Veilleux et al. (1999) are drawn by the blue stars. Note that almost all the 
LINERs with luminous infrared emission are below the boundaries suggested in this letter.
The galaxies selected by the occurrence of core-collapse SNe are plotted by the red circles
(solid for SNe II and open for SNe Ib/c). Among these galaxies, only rare events
are located in the region of the population I LINERs.
}
\end{figure}

\clearpage
\begin{table}
\begin{center}
\caption{COMPARISON OF THE OBSERVATIONAL PROPERTIES BETWEEN THE POPULATION I AND II LINERS}
\begin{tabular}{ccc}
\tableline\tableline
Properties & Population I & Population II\\
\tableline
[\ion{O}{1}]/H$\beta$ & High & Low\\
Broad lines & Yes & No\\
Stellar population & Old & Relatively young \\
Infrared emission & No & Yes\\
Core-collapse SNe & No & Yes\\
\tableline
\end{tabular}
\end{center}
\end{table}

\end{document}